# Nonlinear Charge Transport in InGaAs Nanowires at Terahertz Frequencies

*Rakesh Rana,[1,*] Leila Balaghi,[1,2] Ivan Fotev,[1,2] Harald Schneider,[1] Manfred Helm,[1,2] Emmanouil Dimakis,[1] and Alexej Pashkin[1,*]*

[1]Institute of Ion Beam Physics and Materials Research, Helmholtz-Zentrum Dresden-Rossendorf, 01328 Dresden, Germany

[2]cfaed, Technische Universität Dresden, 01062 Dresden, Germany

* corresponding authors: r.rana@hzdr.de and a.pashkin@hzdr.de

**We probe the electron transport properties in the shell of GaAs/In$_{0.2}$Ga$_{0.8}$As core/shell nanowires at high electric fields using optical pump / THz probe spectroscopy with broadband THz pulses and peak electric fields up to 0.6 MV/cm. The plasmon resonance of the photoexcited charge carriers exhibits a systematic redshift and a suppression of its spectral weight for THz driving fields exceeding 0.4 MV/cm. This behavior is attributed to the intervalley electron scattering resulting in the increase of the average electron effective mass and the corresponding decrease of the electron mobility by about 2 times at the highest fields. We demonstrate that the increase of the effective mass is non-uniform along the nanowires and takes place mainly in their middle part, leading to a spatially inhomogeneous carrier response.**
**Our results quantify the nonlinear transport regime in GaAs-based nanowires and show their high potential for development of nano-devices operating at THz frequencies.**

III-V semiconductor nanowires (NWs) have provided an exceptional platform for high-speed nano(opto-) electronic devices as well as a testbed for various physical phenomena. They allow a large tunability of the bandgap via chemical composition and/or strain and a confined quasi-1D charge transport, while maintaining high carrier mobility and relatively long carrier lifetimes comparable to bulk semiconductors.[1] A wide spectrum of NWs applications have been realized, such as in nano-lasers,[2,3] field effect transistors,[4,5] light emitting diodes,[6,7] solar cells,[8,9] terahertz detectors,[10,11] to name a few.

GaAs has been a popular material choice for NWs,[12,13] which are often doped with suitable impurities[14-17] or coated with an outer shell in order to suppress surface recombination and scattering processes[18-22] or to perform a strain engineering.[23,24] Among various shell materials, In$_x$Ga$_{1-x}$As is quite attractive since its bandgap can be tuned from 1.42 eV to 0.34 eV by increasing the indium content $x$, and it possesses high electron mobility that ranges from 4000 to 18000 cm$^2$V$^{-1}$s$^{-1}$ (for carrier density ~10$^{17}$ cm$^{-3}$) depending on $x$.[25,26] Owing to these favorable properties, In$_x$Ga$_{1-x}$As-based high-electron-mobility transistors (HEMTs) have pushed the operational range into the terahertz regime with power gain frequencies exceeding 1 THz.[27-29] Moreover, implementation of HEMT technology using vertically standing NWs with gate-all-around geometry paves the way for high-performance integrated circuits with unprecedented degree of integration.[5]

In the view of the rapid progress of NW-based electronics and its potential advance above the conventional frequencies of silicon-based devices, there is a strong demand for a study of electron transport in III-V semiconductor NWs at terahertz frequencies in high electric fields. The extremely high operation frequency constrains the length of the transistor channel/base to 20-30 nm[27-29] resulting in the electric fields of 150-250 kV/cm even for an applied voltage of only 0.5 V. However, conventional methods based on direct electric measurements are complicated due to the difficulty of contacting nanometer sized devices and the strong heat dissipation caused by the operation in the high-field regime.

Recently, time-resolved terahertz (THz) spectroscopy has established itself as an efficient tool for probing the charge carrier response in semiconductor NWs.[1] Many important NW parameters such as mobility, scattering rate, carrier concentration and lifetime can be measured accurately using this *contactless* technique. For example, the influence of the NW diameter, the shell thickness and composition, the doping, the surface passivation and other factors on the carrier lifetime and mobility has been extensively studied.[12,13,18,22,30-33] Moreover, the transient character of the applied electric field (a typical THz pulse lasts for about 1 ps) enables measurements in the high-field regime, well beyond the static breakdown limit of GaAs[34] (~ 0.4 MV/cm), without irreversible damage of the sample. In particular, high-field THz experiments in bulk and thin film semiconductors have demonstrated phenomena such as electroluminescence caused by multiplication effects,[35,36] effective mass anisotropy,[37] ballistic electron transport,[38,39] interband tunneling,[40] and field-induced THz bleaching due to intervalley scattering.[41-43]

In this work, we investigate the high-field transport and the plasmonic response of GaAs/In$_{0.2}$Ga$_{0.8}$As core/shell NWs by employing near-infrared pump / THz probe spectroscopy using peak electric fields up to 0.6 MV/cm. We observed a threshold-like behavior of the electron mobility that starts to decrease for peak electric fields above 0.4 MV/cm. This phenomenon is assigned to the increase of the effective mass resulting from a field-driven intervalley scattering in the In$_{0.2}$Ga$_{0.8}$As shell. Remarkably, the mobility drop is accompanied by a noticeable deviation from the linear dependence between the plasmon spectral weight and the square of its frequency indicating an inhomogeneous carrier distribution inside the NWs. Our theoretical modelling demonstrates that this behavior is caused by a non-uniform distribution of the local electric field inside the NWs leading to the intervalley scattering mainly in their middle part.

Undoped GaAs/In$_{0.2}$Ga$_{0.8}$As core/shell NWs with a length of ~2 μm were grown on a Si(111) substrate using molecular beam epitaxy.[23] The 25-nm-thick GaAs cores were covered by an In$_{0.2}$Ga$_{0.8}$As shell with a thickness of 80 nm. The In$_{0.2}$Ga$_{0.8}$As shell and the GaAs core share the same crystallographic orientation (the NW axis is parallel to [111]).[23] After the growth, the vertically-standing NWs were transferred onto a z-cut quartz substrate for THz measurements as shown in Figure 1(a). The NWs are predominantly oriented along the direction of the THz electric field so that about 70% of the NWs experience nearly full strength of the incident field along their axis [see Supporting Information S1]. Previous photoluminescence measurements and the strain analysis demonstrated that the In$_{0.2}$Ga$_{0.8}$As shell is mechanically relaxed and possesses the bandgap of 1.1 eV.[22,23] Owing to a much larger volume as compared to the GaAs core, the In$_{0.2}$Ga$_{0.8}$As shell dominates the transport properties and the plasmonic response of the photoexcited NWs.[22]

The THz response of photoexcited NWs is described with the localized surface plasmon (LSP) model which has been extensively studied in previous works.[1,22] The LSP mode is longitudinal, produced by the THz field component parallel to the NW axes.[1,22] Measured pump-induced change in the optical conductivity $\Delta\sigma(\omega) \approx \sigma(\omega)$ (as the charge carriers are injected optically) can be fitted with a Lorentzian function:

$$\sigma(\omega) = i\frac{ne^2\omega}{m^*(\omega^2 - \omega_0^2 + i\gamma\omega)}, \quad (1)$$

where $n$ is the photo-excited carrier density, $e$ is the elementary charge, $m^*$ is the effective electron mass (the holes are usually neglected because of their significantly larger effective mass and, thus, limited contribution to optical conductivity), $\gamma$ is the scattering rate and $\omega_0$ is the LSP frequency. The last parameter is given by the following relation:

$$\omega_0 = \sqrt{\frac{gne^2}{\epsilon_0\epsilon_w m^*}} = \sqrt{g}\omega_p, \quad (2)$$

where $\varepsilon_w$ is the permittivity of the undoped NW and $g$ is a geometrical factor. The latter parameter is defined by the aspect ratio of the NW and describes the rescaling of the bulk plasma frequency $\omega_p$ via the depolarization field inside the NW that sets the LSP frequency.[1,22] For cylindrical NWs the geometrical factor depends on the aspect ratio of the NW radius and its length.[22]

In contrast to the LSP frequency $\omega_0$, the spectral weight (*SW*) of the LSP mode depends *only* on the ratio $n/m^*$ and is not affected by the aspect ratio of the NW:

$$SW = \int_0^\infty \sigma_1(\omega)d\omega = \frac{\pi ne^2}{2m^*} \quad (3)$$

Thus, the fitting of a plasmon peak with Eq. (1) is performed using three independent parameters: the scattering rate $\gamma$, the LSP frequency $\omega_0$ and the spectral weight *SW*. The latter two values allow us to extract the carrier concentration $n$ and the geometrical factor $g$ as *independent* parameters.

For a preliminary characterization of the NWs in relatively weak THz fields, i.e. the regime of a linear plasmonic response, we utilize a laser system with a repetition rate of 250 kHz. The near-infrared pump with photon energy of 1.55 eV generates electron-hole plasma in the NWs. We record the pump-induced changes in the transmission of the delayed THz probe pulse $\Delta E(t,\tau)$ for different pump-probe delay times $\tau$. The photoconductivity $\sigma(\omega,\tau)$ is calculated using the standard procedure.[1,22] Figure 1(b) shows the LSP mode in the THz photoconductivity for a pump fluence of 45 µJ/cm$^2$ for different delay times. The peak position displays a typical redshift caused by the carrier recombination. The fitting of individual spectra for every delay time using Eq. (1) gives a scattering rate $\gamma/2\pi$ = 1.7±0.3 THz that remains almost constant. The evolution of the LSP frequency and the carrier density shown in Figure 1(c) demonstrates a typical exponential decay. From the mono-exponential fit of $n(\tau)$ we obtain the carrier lifetime of ~70 ps. For the electron effective mass $m^*$ = 0.054$m_e$, the electron mobility in the In$_{0.2}$Ga$_{0.8}$As shell is estimated to be $\mu = e/(m^*\gamma)$ = 2500±200 cm$^2$V$^{-1}$s$^{-1}$.

Knowing the low-field response of the NWs we proceed to the main part of this study - the nonlinear THz measurements in the high-field regime. The experimental scheme is identical to the previous case except the electric field strength of the THz probe pulses, which goes up to 0.6 MV/cm (see Supporting Information S2). The lower panels in Figure 2(a) show the temporal profile of the intense single-cycle THz pulse recorded by electro-optic sampling in a GaP crystal and the corresponding frequency spectrum. The peak electric field reaches 0.6 MV/cm and the frequency spectrum spans up to 7.5 THz with the maximal spectral amplitude around 1.9 THz. In order to achieve the resonant excitation of the LSP mode, we adjust the LSP frequency $\omega_0$ to 1.9 THz by choosing the pump fluence $\Phi = 10$ μJ/cm$^2$ and the pump-probe delay time $\tau = 27$ ps. Figure 2(b) shows that for a moderate probe field of 0.2 MV/cm the LSP mode of the NWs matches the spectral maximum of THz pulse. Using the fitting with Eq. (1), we estimate the carrier density $n = 1.55 \times 10^{17}$ cm$^{-3}$. The relatively long delay of 27 ps is chosen in order to ensure a complete thermalization and an efficient cooling of the photoexcited hot electron distribution via electron-phonon scattering.

Figure 2(b) shows the evolution of the LSP mode upon increasing the driving peak field ($E_{THz}$) from 0.2 to 0.6 MV/cm. Although the photoinduced carrier density is fixed by keeping ($\Phi, \tau$) = constant, the LSP resonance demonstrates a gradual redshift and a suppression of the spectral weight with increasing $E_{THz}$. Figures 2(c) and 2(d) depict $\omega_0$ and $SW$ as functions of $E_{THz}$ obtained from the fitting by Eq. (1). Both parameters remain nearly constant for the fields below ~0.4 MV/cm and start to decrease noticeably for higher fields. According to Eq. (3), the lowering of the spectral weight must be related to the decrease of the ratio $n/m^*$. The most realistic scenario in this case is the increase of the average effective mass $m^*$ caused by intervalley scattering (IVS) of electrons from the $\Gamma$-valley to the $L$- and $X$-valleys of the conduction band as illustrated in Figure 2(a). The acceleration of electrons in the strong THz fields can raise their kinetic energy greater than ~0.38 eV (for the L-valley scattering) initiating the strong IVS via emission or absorption of acoustic or optical phonons. This process has a threshold-like character and it has been previously reported in bulk In$_{0.53}$Ga$_{0.47}$As and GaAs.[41-43]

In order to quantify the impact of the intervalley scattering on the average effective mass $m^*$, we have to know how the electron concentration $n$ depends on the applied THz field. In general, there are two mechanisms that can lead to the generation of additional charge carriers in strong electric fields: impact ionization and interband tunneling. Both of them have been reported in high-field THz experiments on GaAs.[35,36] Impact ionization is known to be particularly efficient since even a single Auger-type interband scattering event per electron leads to a doubling of the electron density.[35] However, our analysis shows that for the NWs where the transport is confined to the [111] direction, the maximum kinetic energy of about 1 eV in the conduction band of In$_{0.2}$Ga$_{0.8}$As is not sufficient for interband scattering [see Supporting Information S3]. Therefore, impact ionization does not take place in the studied NW for any strength of the applied electric field. Furthermore, we estimate the contribution of Zener tunneling to be small compared to the photoinduced density of charge carriers in the NWs [see Supporting Information S3]. Thus, in our analysis we assume $n$ to be independent of $E_{THz}$.

Figure 3(a) shows the field dependence of the average electron effective mass normalized to the nominal mass in the $\Gamma$-valley. It is derived from the spectral weight data (Figure 2(d)) using Eq. (3) and the assumption of the field-independent electron density. Similar to Figures 2(c) and (d), we observe an increase of the effective mass above the threshold field of 0.4 MV/cm due to intervalley transfer and its levelling-off at the level of $m^*/m_\Gamma \approx 2$ for fields above ~0.5 MV/cm. The increase in effective mass is usually attributed to the IVS from $\Gamma$-valley to $L$-valley,[41-42] however, in high fields also the scattering to $X$-valley[44] can play an important role.

The scattering rate $\gamma$ shown in Figure 3(b) also displays a mild decrease above 0.4 MV/cm, which may be another consequence of the side valleys population. Knowing the effective mass and the scattering rate, we can calculate the electron mobility $\mu = e/(m^*\gamma)$ as a function of $E_{THz}$. As it can be seen in Figure 3(c), the mobility for $E_{THz}$ < 0.4 MV/cm is independent of the applied electric field (linear charge transport regime) and matches the value of 2500 cm$^2$V$^{-1}$cm$^{-1}$ measured in the low-field limit. However, the mobility for $E_{THz}$ > 0.4 MV/cm decreases steadily with the electric field (nonlinear charge transport regime), reaching a value of 1000±50 cm$^2$V$^{-1}$s$^{-1}$ for the maximum driving field of 0.6 MV/cm.

Finally, we discuss the scaling between the spectral weight and the LSP frequency. According to Eqs. (2) and (3), $\omega_0^2$ and $SW$ should be proportional to $n/m^*$. However, a comparison between Figures 2(c) and 2(d) shows that the spectral weight decreases faster than the LSP frequency for higher THz fields. This disproportionality becomes obvious in Figure 3(d), which shows $SW$ as a function of $\omega_0^2$. For high THz fields, above the electron transfer threshold, we observe a clear deviation from the linear scaling between both values giving an unequivocal evidence for a change of the geometrical factor $g \propto \omega_0^2/SW$. The dependence of $g$ on the applied THz field is shown in Figure 3(e). The geometric factor exhibits a similar trend as $m^*/m_\Gamma$ in Figure 3(a): for $E_{THz}$ < 0.4 MV/cm the value of $g$ remains nearly the same and it almost doubles for $E_{THz}$ above 0.5 MV/cm. The increase of $g$ points out an increase in the aspect ratio between the radius and the length of the NW. Since it is impossible that the NW radius increases, we assume that the effective length of the NW must be reduced by about 40% as can be estimated from the relation between $g$ and the aspect ratio.[22] In order to understand the origin of this behavior we have performed COMSOL simulations of the electric field distribution and the THz response of a single NW [see Supporting Information S4]. First, we assume the low-field limit with electrons in the $\Gamma$-valley uniformly distributed along the NW. Figure 4(a) shows the calculated real part of the optical conductivity (black symbols) with a single peak at 1.75 THz, which corresponds to the low-field plasmon mode LSP0. Figure 4(b) depicts the inhomogeneous field distribution across the NW at the frequency of 1.9 THz, which corresponds to the spectral maximum of the probe pulse. As expected, the field is almost completely screened at the ends of the NW by the plasmon charge accumulated at the NW tips. On the other hand, in the middle of the NW the screening is negligible, and the field is even slightly enhanced as compared to the external field due to the LSP resonance. Obviously, in the high-field regime, the invervalley scattering should be dominant in the middle part of the NW, whereas the parts near the NW ends should not be affected. Thus, the effective mass increase should occur non-uniformly along the NW. Taking into account the relatively steep threshold-like character of the mass dependence on $E_{THz}$, we model the high-field response of the NW by splitting it into three parts: the middle part with a length of 1.2 μm (60% of the total NW length) where the mass enhancement takes place ($m^*/m_\Gamma = 2$) and the two ends with a length of 0.4 μm each where the effective mass remains unchanged ($m^*/m_\Gamma = 1$). Figure 4(b) shows the calculated optical conductivity for this model (red symbols), which demonstrates two peaks denoted as LSP1 and LSP2. The dominant LSP1 mode redshifted with respect to the LSP0 down to 1.3 THz corresponds to the plasmon resonance in the middle part of the NW as proved by the field distribution shown in the same figure. Namely, this mode with the enhanced $g$ due to the shorter length of the active region is observed in our experiment. The high-frequency mode LSP2 at 3.4 THz corresponds to the plasmonic resonance in the NW ends as can be seen from the field distribution shown in the inset of Figure 4(a). It is blue-shifted with respect to the LSP0 due to a larger geometrical factor – the end parts are 5 times shorter than the whole NW. As a result the LSP2 mode is located outside of the experimentally accessible frequency region and also possess rather small spectral weight due to the relatively small portion of the charge carriers contributing to this mode.

Overall, the simulation demonstrates that even the simplistic model of inhomogeneous effective mass distribution along the NW is able to capture the main features observed in the experiement. More realistic modelling has to consider a continous distribution of the effective mass along the NW, which depends on the local field strength. Moreover, the model should be self-consistent since any change in the electron mass distribution also changes the field distribution inside the NW as it can be clearly seen by comparing Figures 4(b) and 4(c). Although a full modelling is beyond the scope of the present work, we anticipate that the self-consistent calculations may explain the saturation of the geometrical factor at high fields that is observed in our experiment (see Figure 3(e). The region where the IVS dominates in the high-field regime does not expand further for even higher fields since the field distribution becomes localized around the active volume of the LSP1 as illustrated in Figure 4(c).

In conclusion, we have investigated the charge carrier response in $GaAs/In_{0.20}Ga_{0.80}As$ core/shell NWs to THz fields with amplitudes up to 0.6 MV/cm. Field-driven electron transfer from the $\Gamma$-valley to the side valleys of the conduction band in the $In_{0.20}Ga_{0.80}As$ shell sets in at 0.4 MV/cm, leading to an increase of the effective mass by a factor of 2 for the highest applied fields. Furthermore, we demonstrate that the intervalley scattering occurs mainly in the middle part of the NW resulting in a spatially inhomogeneous effective mass distribution. The high electron mobility and the suppressed impact ionization in sizable electric fields prove a large potential of epitaxial NWs for applications in terahertz band nanoelectronics.

The authors thank Jiang Li and J. Michael Klopf for the help with the calibration of the THz electric field strength.

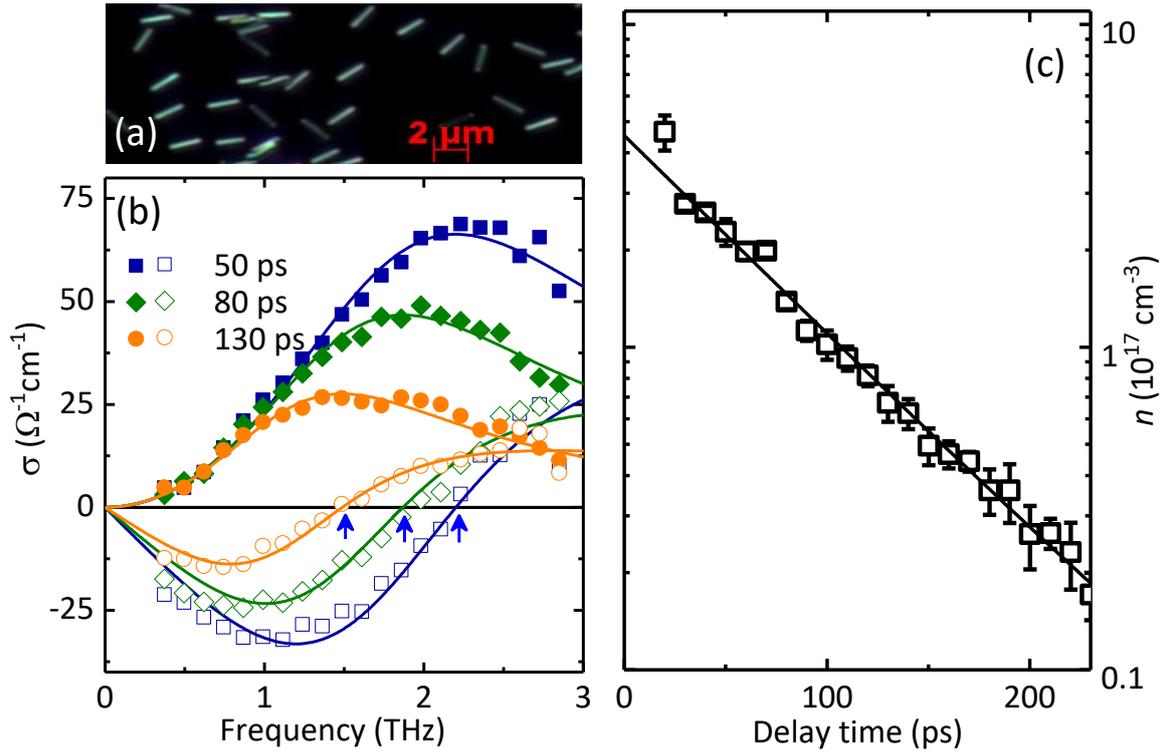

**Figure 1**. (a) Optical microscopy image taken in the dark-field mode for the GaAs/In$_{0.2}$Ga$_{0.8}$As core/shell nanowires sample transferred onto a *z*-cut quartz. (b) Frequency dependent photoconductivity ($\sigma$) for different pump probe delay times at a pump fluence ($\Phi$) of 45 µJ/cm$^2$, closed and open symbols are the real and imaginary part of $\sigma$ and the solid lines are fits with the LSP model. (c) Temporal evolution of the carrier concentration *n* for different pump-probe delay times and a mono-exponential fit (black line).

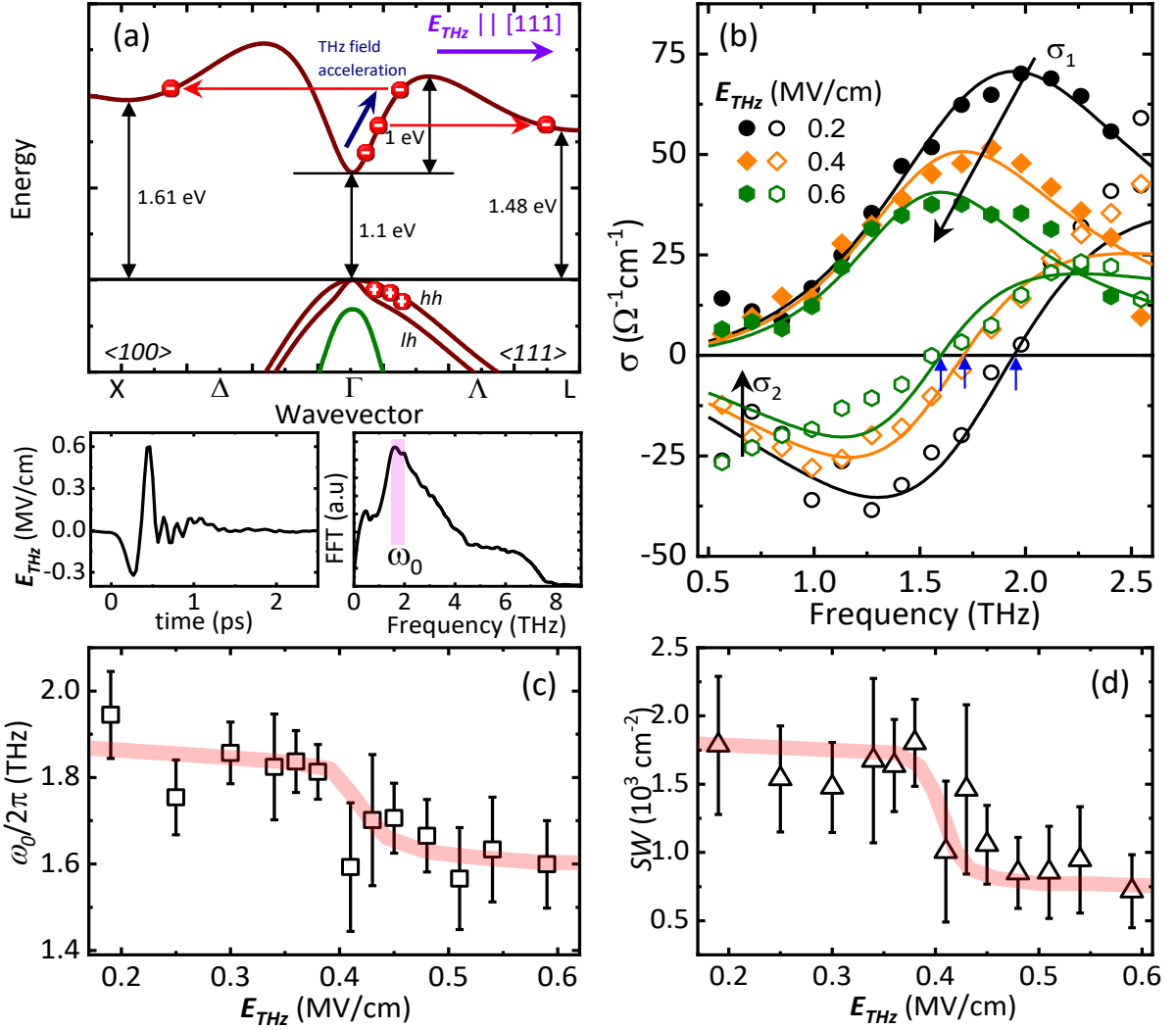

**Figure 2**. (a) Schematic band structure of the In$_{0.2}$Ga$_{0.8}$As shell. The values of the band parameters are estimated based on data from Ref. 45. The blue arrow marks the acceleration of electrons by the THz field applied in the [111] direction of the NW axis. The red arrows designate intervalley scattering processes. The bottom left and right panels shows the THz probe pulse and its spectrum, with the pink area denoting the plasmon frequency $\omega_0$ in the low-field limit. (b) Real and imaginary part of photoconductivity ($\sigma$) for several THz field strengths depicted by closed and open symbols, respectively. The pump fluence is 10 μJ/cm$^2$ and the pump-probe delay time is 27 ps. Solid lines are fits by the LSP model. (c) The plasmon frequency and (d) the spectral weight as functions of the terahertz field ($E_{THz}$).

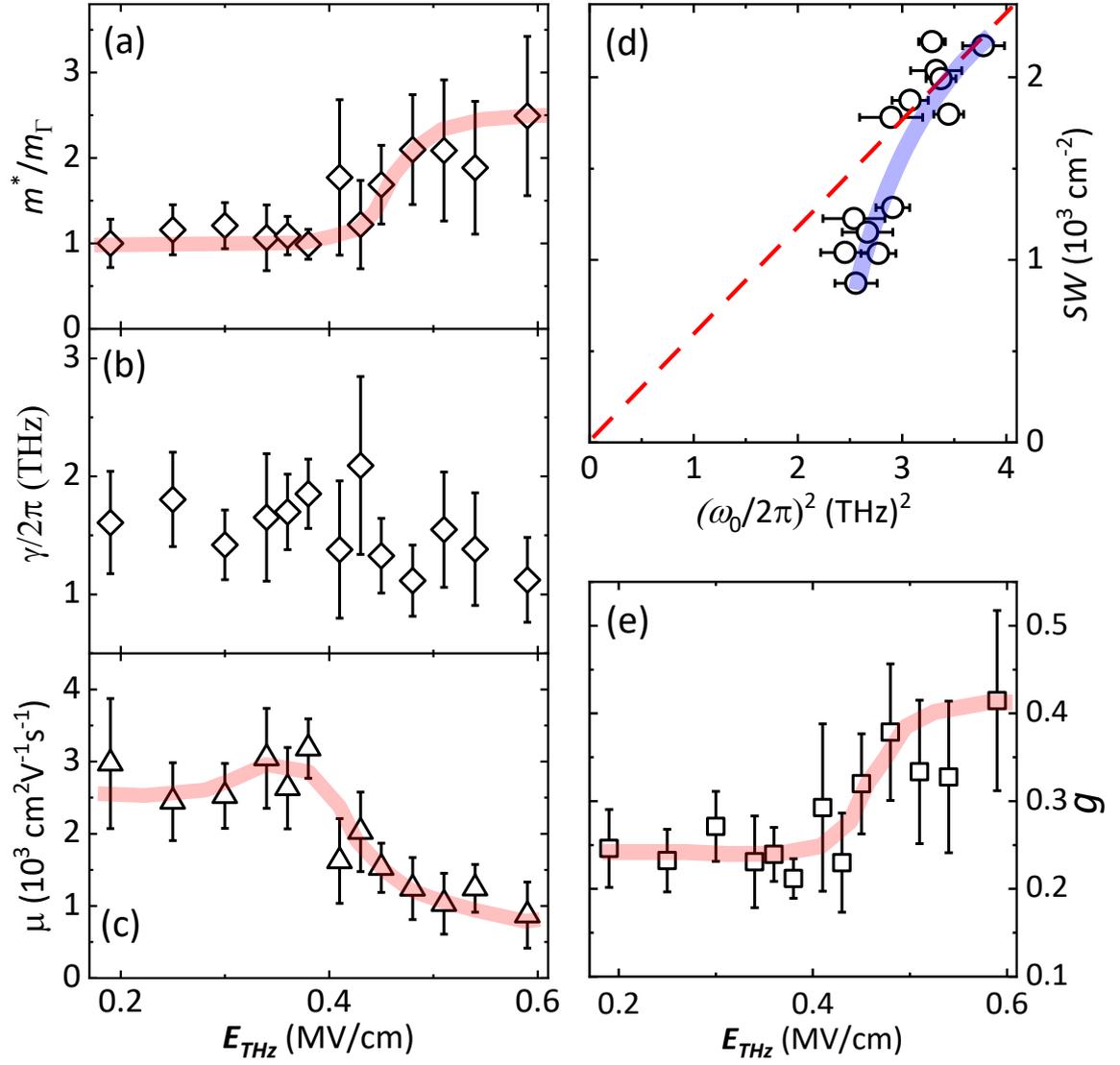

**Figure 3.** (a) Normalized effective mass $m^*/m_\Gamma$, (b) scattering rate $\gamma/2\pi$ and (c) mobility $\mu$ as a function of the THz field ($E_{THz}$). (d) Spectral weight versus square of the LSP frequency. (e) Geometrical factor $g$ as a function of $E_{THz}$. The thick lines are guides to the eye.

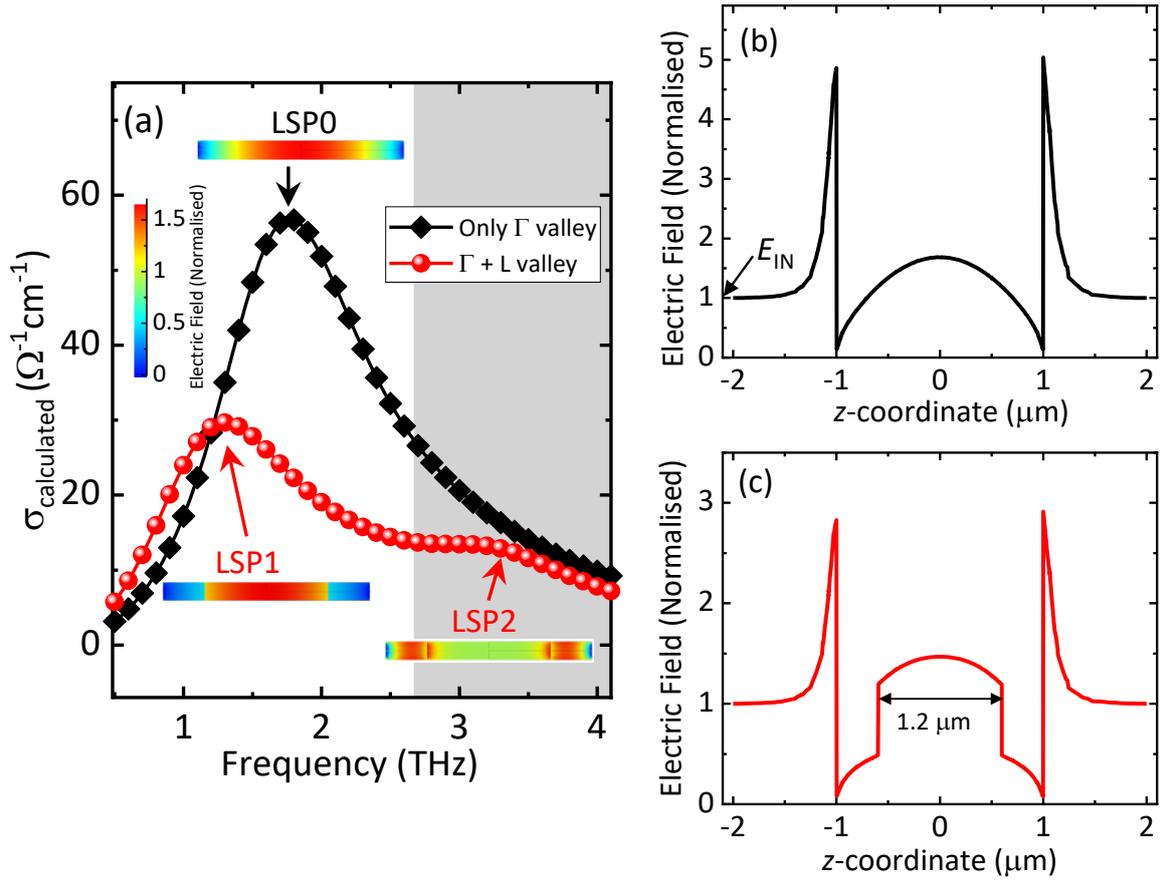

**Figure 4**. (a) Real part of the nanowire conductivity ($\sigma_{calculated}$) simulated using the *COMSOL Multiphysics* software. The black dots correspond to the linear regime ($m^*/m_\Gamma = 1$) with a single resonance LSP0 at 1.75 THz. The red dots correspond to the nonlinear high-field regime ($m^*/m_\Gamma = 2$ in the central part of the NW and $m^*/m_\Gamma = 1$ at its edges) and display two resonances: LSP1 at 1.3 THz and LSP2 at 3.4 THz. The field distributions inside the NWs for each resonance are shown in the inset. The shaded area marks the frequencies beyond the range covered by the experiment. (b) and (c) show the normalized electric field component along the NW axis at the central frequency of the THz probe field (1.9 THz) for the linear and nonlinear regimes, respectively.

# Supporting Information

**S1 Directionality analysis of nanowires transferred onto a Quartz substrate**

The free-standing GaAs/In$_{0.2}$Ga$_{0.8}$As core/shell nanowires (NWs) with 2 μm length are grown using MBE on a Si(111) oriented substrate. For terahertz (THz) measurements, NWs are mechanically transferred onto a z-cut quartz substrate by sliding the surface of the Si substrate with the NWs along the quartz surface. In this way, the sliding direction defines the dominating orientation of the NWs on the quartz substrate. However, the transferred NWs are perfectly oriented and demonstrate a spread of orientations as shown in figure S1(a). In order to quantify the degree of the orientation we have used the "Directionality" plug-in of the Fiji software. The resulting orientation histogram is shown in figure S1(b). The zero of the angle φ is defined as a horizontal direction in figure S1(a), it corresponds to the sliding direction during the NW transfer. As the THz field is horizontal, a NW oriented at the angle φ experiences its projection on the NW axis that is proportional to cos φ. Thus, the angular spread of ±25º corresponds to the electric fields exceeding 90% of the incident field, i.e., cos φ > 0.9. This region is pink shaded in figures S1(b,c) and it contains nearly 70 % of the NWs in the histogram. Moreover, the effective contribution of these NWs to the plasmon spectral weight is even larger since our experiment detects only the horizontal component of the THz-induced polarization along the NW axis. Thus, a NW contribution to the optical conductivity is proportional to cos$^2$φ. The relative spectral weight contribution of NWs in the pink area can be calculated as $\int_{-25}^{25} A(\varphi) \cos^2\varphi \, d\varphi / \int_{-90}^{90} A(\varphi) \cos^2\varphi \, d\varphi \approx 0.8$. This means that 70% of all NWs experience more than 90% of the maximal THz field and they contribute about 80% to the total plasmonic response. Therefore, the orientation degree of our sample is sufficient for the observation of the field-driven threshold behavior although its onset is expected to be smeared due to certain portion of less oriented NWs.

**S2. Optical pump / terahertz probe (high-field) measurement setup**:

For high-field nonlinear measurements we utilize a Ti:sapphire regenerative amplifier delivering 6 mJ pulses with 35 fs FWHM duration at a 1 kHz repetition rate with the central wavelength of ~ 800 nm (1.55 eV). The output beam is split 50:50. One half is used to pump the optical parametric amplifier (OPA) and the other half is further split using a 90:10 beam splitter. The stronger portion of the latter serves as an optical pump for the nanowire sample and the weaker portion is used for electro-optic sampling of the THz pulses using 1-mm-thick (110) ZnTe or 0.4-mm-thick (110) GaP detector crystals for the pump-probe measurement and the pulse characterization, respectively.

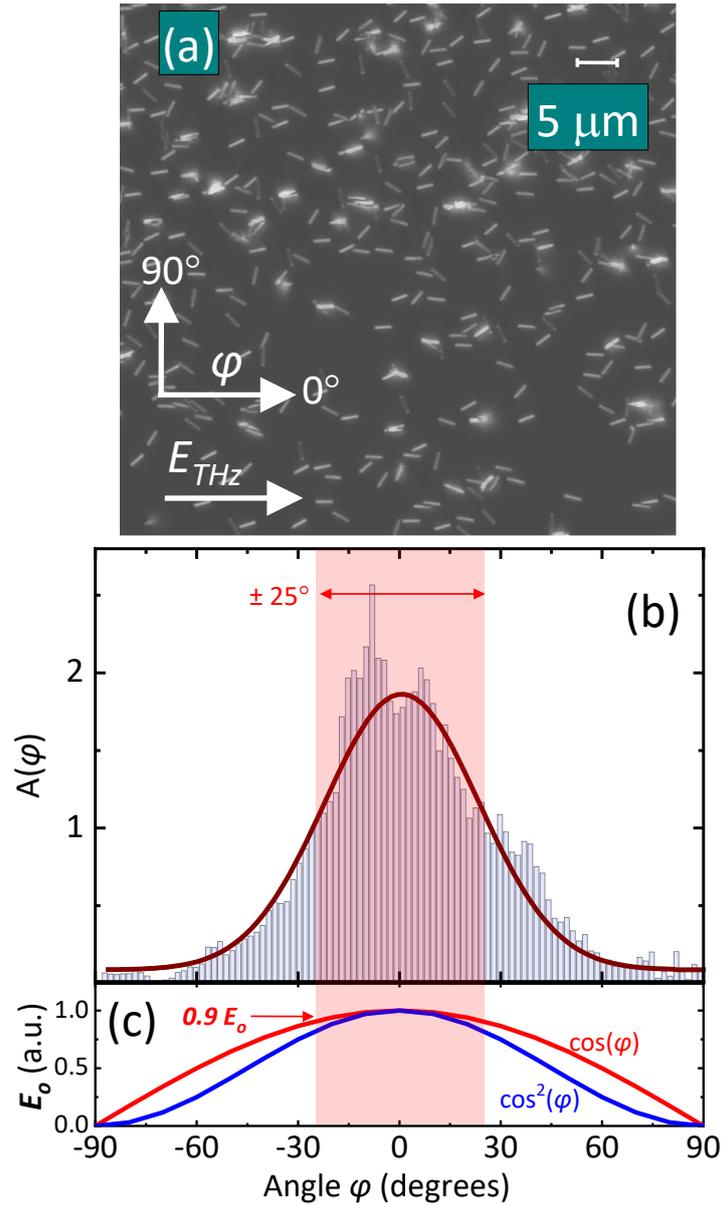

**Figure S1.** (a) Optical microscopy image taken in the dark-field mode of the GaAs/In$_{0.2}$Ga$_{0.8}$As NWs on a quartz substrate. The terahertz field (**E$_{THz}$**) is applied at an angle φ = 0°; (b) The histogram of the angular NW distribution, the brown line is the Gaussian fit; (c) The projection of the electric field *E$_0$* on the NW axis as a function of the NW angle φ (red curve) and the corresponding contribution of the NWs to the spectral weight (blue curve).

The OPA delivers intense near-infrared pulses with the energy up to 0.4 mJ at the center wavelength of 1425 nm. The high-field single-cycle terahertz pulses with max. energy of 0.37 µJ are generated via optical rectification of the collimated OPA beam in a DSTMS crystal (thickness 400 µm). To filter out the residual OPA beam after the DSTMS crystal we used a double side polished Ge wafer placed at the Brewster's angle and a black polyethylene foil. The pulse energy was measured using a **THz 20** pyroelectric detector (SLT Sensor- und Lasertechnik GmbH), which was calibrated around 1.4 THz at the Physikalisch-Technische Bundesanstalt (PTB). The THz radiation is focused on the sample by an off-axis parabolic mirror with the effective focal length of 50 mm to a nearly diffraction-limited spot with a FWHM diameter of ~650 µm. The peak electric field of the THz transient reaching 0.6 MV/cm was estimated based on the THz fluence and the electro-optically measured temporal profile depicted in Figure 2(b) of the main paper.

The 800 nm wavelength pump beam is focused on the sample to a FWHM diameter of approximately 1.5 mm. To avoid any nonlinearity in the electro-optic detection we place a pair of wire grid polarizers after the nanowire sample to attenuate the intense THz signal on the ZnTe detector. All measurements were performed at room temperature in a box purged with a dry nitrogen gas.

## S3. Impact ionization and inter-band Zener tunneling

Here we discuss the contributions of the impact ionization and the inter-band tunneling on the carrier concentration in the NWs.

Impact ionization occurs when an electron in the conduction band gains enough kinetic energy to excite a new electron-hole (*e-h*) pair via Auger interband scattering. Since momentum as well as energy conservation must be fulfilled simultaneously, the threshold electron energy ($E_{th}$) for impact ionization is somewhat larger than the bandgap. It is given by the equation $E_{th} = E_g(2m_e + m_{hh})/(m_e + m_{hh})$, where $E_g$ is the bandgap and $m_e$ and $m_{hh}$ are the effective masses of electrons and heavy holes, respectively.[1,2] For our case of the relaxed $In_{0.2}Ga_{0.8}As$ shell, $E_{th} \approx 1.25$ eV. This value is well above the maximum kinetic energy of about 1 eV along the $\Gamma$-$L$ line of the $In_{0.2}Ga_{0.8}As$ band structure.[2] Thus, impact ionization is completely suppressed in (111)-oriented NWs due to the confined character of the electron transport.

Another process that can lead to an increase of the electron concentration in high electric fields is interband tunneling or multiphoton absorption. The Keldysh parameter $\gamma_K$ can be used to determine which of these limiting cases is applicable.[3] It is defined as $\gamma_K = \omega \sqrt{2m^*E_g}/eE_o$, where $\omega$ is the field frequency, $m^*$ is the effective electron mass and $E_o$ is the amplitude of the driving THz field. In our case, assuming the frequency to be equal to the LSP frequency of 1.9 THz, for a field amplitude of 0.5 MV/cm we get $\gamma_K = 0.19 \ll 1$ indicating the interband tunneling regime for the relevant range of the field amplitudes used in our study. Therefore, the Zener tunneling rate can be calculated using the model developed by Kane[4,5]

$$r_Z(t) = \frac{e^2 E_0^2 m_r^{1/2}}{18\pi\hbar^2 E_g^{1/2}} \exp\left(\frac{-\pi m_r^{1/2} E_g^{3/2}}{2\hbar e E_o}\right).$$

Here the reduced mass $m_r = 0.016 m_e$ accounts for transitions from the light hole initial states to the conduction band. The tunneling rate ($r_Z$) as a function of THz field is shown in figure S2. Here we can neglect the contribution from the heavy holes as it has more than three orders in magnitude less contribution than light holes for peak fields upto 0.6 MV/cm. The additional density of carriers induced by the high-field THz pulses can be estimated by integrating the tunneling rate over the temporal profile of the THz transient

$$\Delta n = \int r_z(E_0(t))dt.$$

The main contribution to the tunneling comes from the half-cycle of the terahertz pulse with the highest electric field. For the peak field of 0.6 MV/cm, the estimated increase in the carrier concentration $\Delta n \approx 2.5 \times 10^{12}$ cm$^{-3}$, which is negligible compared to the initial photo-generated carrier concentration of $1.55 \times 10^{17}$ cm$^{-3}$. Thus, the contribution of interband tunneling can be neglected for our analysis.

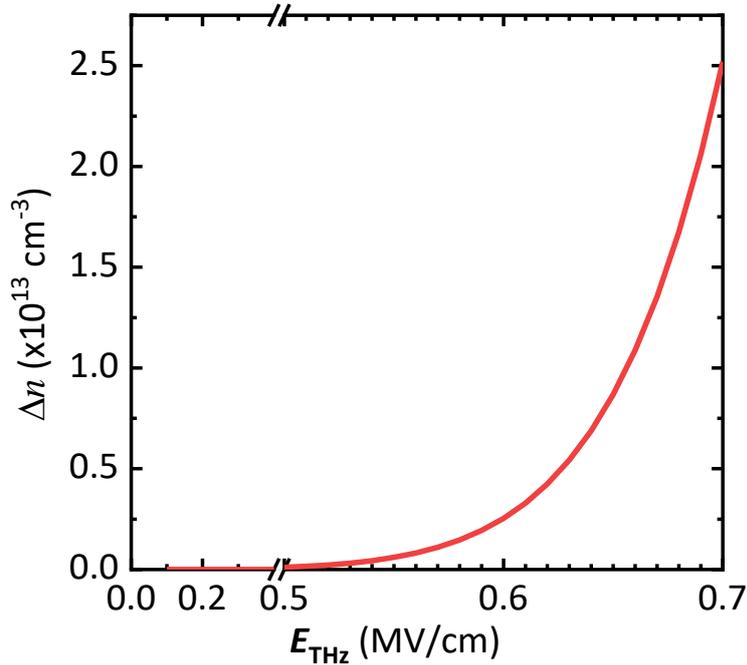

**Figure S2.** (a) Additional carrier density produced by the interband tunneling as a function of the peak THz field ($E_{THz}$).

## S4. Numerical simulation of the plasmon redshift

We analyze numerically (finite element analysis) the response of a photoexcited NW to an external electric field $E_{THz}$ using the *COMSOL Multiphysics* software. The NW is modeled as an In$_{0.2}$Ga$_{0.8}$As cylinder, surrounded by air. The simulation is carried out for a monochromatic wave with the frequency varying from 0.3 to 5 THz. The software solves numerically the wave equation

$$\nabla \times (\nabla \times E(r,\omega)) - \frac{\omega^2}{c^2}\varepsilon(r,\omega)E(r,\omega) = 0, \quad (1)$$

where $c$ is the speed of light in vacuum and $\varepsilon(r,\omega)$ is the dielectric function of the NW material or the surrounding air. Since the THz spot size is much larger compared to a single NW, the Fourier components of $E_{THz}$ were assumed to be background plane waves: $E_b=\exp(i\omega t-i\omega x/c)e_z$. In order to examine the longitudinal LSP mode, we orient the NW in the $z$ direction, i.e. parallel to $E_{THz}$. Then the equation (1) was solved for the total electric field $E=E_b+E_{relative}$ (scattered field formulation).

The dielectric function of the photoexcited $In_{0.2}Ga_{0.8}As$ material is represented using the Drude model[6] (adapted to the sign convention used in *COMSOL*):

$$\varepsilon(\omega) = \varepsilon_w \left[1 - \frac{\omega_p^2}{\omega(\omega-i\gamma)}\right], \qquad (2)$$

where $\omega_p = \sqrt{\frac{ne^2}{\epsilon_0 \epsilon_w m^*}}$ is the bulk plasma frequency. The concentration of photoexcited electrons was assumed to be $n = 1.55\times10^{17}$ cm$^{-3}$ and static dielectric permittivity of $In_{0.2}Ga_{0.8}As$ $\varepsilon_w$=13.23.

First, we consider the linear case, when all electrons are in the $\Gamma$-valley. Therefore, we describe the entire NW with a spatially homogeneous dielectric function $\varepsilon_\Gamma(\omega)$, where $m^* = m_\Gamma = 0.054m_e$, and solve the model for total electric field $E$.

For describing the nonlinear redshift of the LSP frequency, we divide the NW in three parts. The two parts at the NW tips (each 400 nm long) preserve the light $\Gamma$-valley electrons. For these regions, we use the same dielectric function $\varepsilon_\Gamma(\omega)$ as in the first case. The middle part, 1200 nm long, contains the mixture of $\Gamma$- and $L$-valley electrons so that the average effective mass is doubled $m^* = 2m_\Gamma = 0.108m_e$. These assumptions are based on the experimental results for the enhancement of the effective mass ($m^*/m_\Gamma \approx 2$) and the increase in the geometric factor from 0.228 to ~0.4. The latter change implies the decrease in the NW's length by 40% based on analytical estimations by Fotev et al.[7]

In order to calculate the optical conductivity spectra, we determine the total dissipated power inside the NW $P(\omega)$ as a function of frequency. Then the real part of the optical conductivity shown in Figure 4(a) is given by the formula $\sigma_{calculated}(\omega) = 2P(\omega)/E^2V$, where $V$ is the volume of the NW and $E$ is the amplitude of the incident wave (set to 1 V/m in our simulations).